\documentclass[twocolumn,showpacs,preprintnumbers,amsmath,amssymb]{revtex4}
\usepackage{graphicx}
\usepackage{dcolumn}
\usepackage{bm}
\begin{document}
\title{Diffusion-controlled death of $A$-particle and $B$-particle islands at
propagation of the sharp annihilation front
$A + B \rightarrow 0$}
\author{Boris M. Shipilevsky}
\affiliation{Institute of Solid State Physics, Chernogolovka,
Moscow district, 142432, Russia}
\begin{abstract}
We consider the problem of diffusion-controlled evolution of the
system $A$-particle island - $B$-particle island at propagation of
the sharp annihilation front $A+B\to 0$. We show that this general
problem, which includes as particular cases the sea-sea and the
island-sea problems, demonstrates rich dynamical behavior from
self-accelerating collapse of one of the islands to synchronous
exponential relaxation of the both islands. We find a universal
asymptotic regime of the sharp front propagation and reveal limits
of its applicability for the cases of mean-field and fluctuation
fronts.
\end{abstract}
\pacs{82.20.-w, 05.40.-a}
\maketitle

For the last decades the reaction-diffusion system $A +
B\rightarrow 0$, where unlike species $A$ and $B$ diffuse and
annihilate in a $d$-dimensional medium, has acquired the status of
one of the most popular objects of research. This attractively
simple system, depending on the initial conditions and on the
interpretation of $A$ and $B$ (chemical reagents, quasiparticles,
topological defects, etc.), provides a model for a broad spectrum
of problems \cite{kotr}, \cite{taur}. A crucial feature of many
such problems is the dynamical {\it reaction front} - a localized
reaction zone which propagates between domains of unlike species.

The simplest model of a reaction front, introduced almost two
decades ago by Galfi and Racz (GR) \cite{gal}, is a
quasi-one-dimensional model for two initially separated reactants
which are uniformly distributed on the left side ($x<0$) and on
the right side ($x>0$) of the initial boundary. Taking the
reaction rate in the mean-field form $R(x,t)=ka(x,t)b(x,t)$, GR
discovered that in the long-time limit $kt\to\infty$ the reaction
profile $R(x,t)$ acquires the universal scaling form
\begin{eqnarray}
R=R_{f}{\cal Q}\left(\frac{x-x_{f}}{w}\right),
\end{eqnarray}
where $x_{f}\propto t^{1/2}$ denotes the position of the reaction
front center, $R_{f}\propto t^{-\beta}$ is the height, and $
w\propto t^{\alpha}$ is the width of the reaction zone.
Subsequently, it has been shown \cite{cor1}-\cite{bar} that the
mean field approximation can be adopted at $d \geq d_{c}=2$,
whereas in 1D systems fluctuations play the dominant role.
Nevertheless, the scaling law (1) takes place at all dimensions
with $\alpha=1/6$ at $d \geq d_{c}=2$ and $\alpha=1/4$ at $d=1$,
so that at any $d$ the system demonstrates a remarkable property
of the effective "{\it dynamical repulsion}" of $A$ and $B$: on
the diffusion length scale $L_{D}\propto t^{1/2}$ the width of the
reaction front asymptotically contracts unlimitedly: $w/L_{D}\to 0
\quad {\rm as} \quad t\to\infty.$ Based on this property a general
concept of the front dynamics, the quasistatic approxmation (QSA),
has been developed \cite{cor1}, \cite{lee}, \cite{bar}, \cite{ben}
which consists in the assumption that for sufficiently long times
the kinetics of the front is governed by two characteristic time
scales. One time scale $t_{J}=-(d\ln J/dt)^{-1}$ controls the rate
of change in the diffusive current $J=J_{A}=|J_{B}|$ of particles
arriving at the reaction zone. The second time scale $t_{f}\propto
w^{2}/D$ is the equilibration time of the reaction front. Assuming
that $t_{f}/t_{J}\ll 1$ from the QSA in the mean-field case with
$D_{A,B}=D$ it follows that \cite{cor1}, \cite{lee}, \cite{ben}
\begin{eqnarray}
R_{f}\sim J/w, \quad w\sim (D^{2}/Jk)^{1/3},
\end{eqnarray}
whereas in the 1D case $w$ acquires the $k$-independent form
$w\sim (D/J)^{1/2}$ \cite{cor1}, \cite{lee}. On the basis of the
QSA a general description of spatiotemporal behavior of the system
$A+B\to 0$ has been obtained for arbitrary nonzero diffusivities
\cite{koza2} which was then generalized to anomalous diffusion
\cite{kat1}, diffusion in disordered systems \cite{hec}, diffusion
in systems with inhomogeneous initial conditions \cite{ran}, and
to several more complex reactions. Following the simplest GR model
\cite{gal} the main attention has been traditionally focused on
the systems with $A$ and $B$ domains having an unlimited
extension, i.e., with {\it unlimited number} of $A's$ and $B's$
particles, where asymptotically the stage of monotonous
quasistatic front propagation is always reached: $t_{f}/t_{J}\to 0
\quad {\rm as} \quad t\to\infty.$

Recently, in the work \cite{ship} a {\it new line} in the study of
the $A+B\to 0$ dynamics has been developed under the assumption
that the particle number of one of the species  is {\it finite},
i.e., an $A$ particle island is surrounded by the uniform sea of
particles $B$. It has been established that at sufficiently large
initial number of $A$ particles $N_{0}$ and a sufficiently high
reaction rate constant $k$ the death of the majority of island
particles $N(t)$ proceeds in the  {\it universal scaling regime}
$N=N_{0}{\cal G}(t/t_{c})$, where $t_{c}\propto N_{0}^{2}$ is the
lifetime of the island in the limit $k,N_{0}\to\infty$. It has
been shown that while dying, the island first expands to a certain
maximal amplitude  $x_{f}^{M}\propto N_{0}$ and then begins to
contract by the law $x_{f}=x_{f}^{M}\zeta_{f}(t/t_{c})$ so that on
reaching $x_{f}^{M}$ (the turning point of the front)
\begin{eqnarray}
t_{M}/t_{c}= 1/e, \quad N_{M}/N_{0}=0.19886...
\end{eqnarray}
and, therefore, irrespective of the initial particle number and
dimensionality of the system $\approx 4/5$ of the particles die at
the stage of the island expansion and the remaining $\approx 1/5$
at the stage of its subsequent contraction.

In this Rapid Communication we consider a much more general
problem of the $A+B\to 0$ annihilation dynamics with the initially
separated reactants under the assumption that the particle number
of the {\it both species} is finite. More precisely, we consider
the problem on diffusion-controlled death of $A$-particle and
$B$-particle islands at propagation of the sharp annihilation
front $A+B\to 0$. We show that this island-island (II) problem, of
which particular cases are the GR sea-sea (SS) problem and the
island-sea (IS) problem \cite{ship}, exhibits rich dynamical
behavior and we reveal its most essential features.

Let in the interval $x\in [0,L]$ particles $A$ with concentration
$a_{0}$ and particles $B$ with concentration $b_{0}$ be initially
uniformly distributed in the islands $x\in [0,\ell)$ and $x\in
(\ell,L]$, respectively. Particles $A$ and $B$ diffuse with
diffusion constants $D_{A}$ and $D_{B}$ and when meeting they
annihilate $A+B\to 0$ with a reaction constant $k$. We will
assume, as usually, that concentrations $a(x,t), b(x,t)$ change
only in one direction (flat front) and we will consider that the
boundaries $x=0,L$ are impenetrable. Thus, our effectively one
dimensional problem is reduced to the solution of the problem
\begin{eqnarray}
\partial a/\partial t = D_{A}\nabla^{2} a - R, \quad
\partial b/\partial t = D_{B}\nabla^{2} b - R
\end{eqnarray}
in the interval $x\in [0, L]$ at the initial conditions $
a(x,0)=a_{0}\theta(\ell-x), \quad b(x,0)=b_{0}\theta(x-\ell)$ and
the boundary conditions $\nabla (a, b)\mid_{x=0,L}=0$ where
$\theta(x)$ is the Heaviside step function. To simplify the
problem essentially we will assume $D_{A}=D_{B}=D$. Then, by
measuring the length, time, and concentration in units of $L$,
$L^{2}/D$, and $b_{0}$, respectively, i.e. assuming $L=D=b_{0}=1$,
and defining the ratio of initial concentrations $a_{0}/b_{0}=r$
and the ratio $\ell/L=q$, we come from (4) to the simple diffusion
equation for the difference concentration $s=a-b$
\begin{eqnarray}
\partial s/\partial t = \nabla^{2} s,
\end{eqnarray}
in the interval $x\in [0, 1]$ at the initial conditions
\begin{eqnarray}
s_{0}(x\in [0,q))=r, \quad s_{0}(x\in (q,1])=-1,
\end{eqnarray}
with the boundary conditions
\begin{eqnarray} \nabla
s\mid_{x=0,1}=0.
\end{eqnarray}
According to the QSA for large $k\to\infty$ at times $t\propto
k^{-1}\to 0$ there forms a sharp reaction front $w/x_{f}\to 0$ so
that the solution $s(x,t)$ defines the law of its propagation
$s(x_{f},t)=0$ and the evolution of particle distributions $a=s
(x<x_{f})$ and $b=|s| (x>x_{f})$. In the limits sea-sea
\cite{gal}{($\ell\to\infty, L\to\infty$) or island-sea \cite{ship}
($\ell$ finite, $L\to\infty$) the corresponding solutions $s_{\rm
SS}(x,t)$ and $s_{\rm IS}(x,t)$ describe the initial stages of the
system's evolution at times $\sqrt{t}\ll q, 1-q$ and $q \ll
\sqrt{t}\ll 1$, respectively. The general solution to Eqs.(5)-(7)
for arbitrary $r$, $q$ and $t$ has the form
\begin{eqnarray}
s(x, t)=\Delta+\sum_{n=1}^{\infty}A_{n}(r,q)\cos(n\pi
x)e^{-n^{2}\pi^{2}t},
\end{eqnarray}
where coefficients $A_{n}(r, q)=2(r+1)\sin(n\pi q)/n\pi$ and
$\Delta(r,q) = N_{A}-N_{B}= rq - (1-q)$ is the difference of the
reduced number of $A$ and $B$ particles which remains constant. At
$t> 1/\pi^{2}$ we find
\begin{eqnarray}
s=\Delta + A_{1}(r, q)\cos(\pi x)e^{-\pi^{2}t}+\cdots.
\end{eqnarray}
Taking $s(x_{f},t)=0$ we obtain from (9) the law of the front
motion
\begin{eqnarray} \cos(\pi
x_{f})={\cal C}e^{\pi^{2}t}+\cdots,
\end{eqnarray}
where coefficient ${\cal C}$ can be represented in the form
\begin{eqnarray}
{\cal C}= -\Delta/A_{1} =
q(r_{\star}-r)/A_{1}=(q_{\star}-q)/q_{\star} A_{1},
\end{eqnarray}
where $q_{\star}=1/(r+1)$ and $r_{\star}=(1-q)/q$ are the critical
values of $q, r$ at which ${\cal C}$ reverses its sign. From
Eq.(10) it follows that at $|{\cal C}|< 1/e$ and $r\neq r_{\star},
q\neq q_{\star}$, when the ratio of the initial particle numbers
\begin{eqnarray}
\rho=\frac{N_{A0}}{N_{B0}}=\frac{r}{r_{\star}}=\frac{(1-q_{\star})q}{(1-q)q_{\star}}\neq
1
\end{eqnarray}
the front  $x_{f}(t)$ moves either towards the boundary $x=0$
($\rho < 1$) or towards the boundary $x=1$ ($\rho
> 1$) so that in the limit $k\to\infty$ the island of a smaller particle number
($A$ or $B$, respectively) dies within a finite time
\begin{eqnarray}
t_{c}=(1/\pi^{2})|\ln|{\cal C}||.
\end{eqnarray}
From Eqs. (10) and (13) we obtain
\begin{eqnarray}
x_{f}=(1/\pi)\arccos(\pm e^{\pi^{2}(t-t_{c})}),
\end{eqnarray}
(here and in what follows the upper sign corresponds to $\rho < 1$
and the lower sign corresponds to $\rho> 1$) whence for the front
velocity $v_{f}=\dot{x}_{f}$ we find
\begin{eqnarray}
v_{f} = -\pi \cot(\pi x_{f})=\mp
\pi/(\sqrt{e^{2\pi^{2}(t_{c}-t)}-1}).
\end{eqnarray}
Making use then (13), for the distribution of particles
($a=s(x<x_{f}), b=|s|(x>x_{f})$ \cite{ship}) at $\rho\neq 1$ we
obtain
\begin{eqnarray}
s=\Delta(1\mp\cos(\pi x)e^{\pi^{2}(t_{c}-t)})+\cdots.
\end{eqnarray}
Thus from the condition $N_{A}=\int_{0}^{x_{f}} s dx =
N_{B}+\Delta$ we find the laws of decay of the $A$ and $B$
particle number
\begin{eqnarray}
N_{A}= (|\Delta|/\pi)(\sqrt{e^{2\pi^{2}(t_{c}-t)}-1}\mp \pi x_{f})
\end{eqnarray}
and then we derive finally the diffusive boundary current in the
vicinity of the front
\begin{eqnarray}
J= -\partial s/\partial x\mid_{x=x_{f}}= \pi
|\Delta|\sqrt{e^{2\pi^{2}(t_{c}-t)}-1},
\end{eqnarray}
which according to (2) defines evolution of the amplitude
$R_{f}(t)$ and of the width of the front $w(t)$.

From Eqs. (13)-(18) we immediately come to the following important
conclusions: for arbitrary $r$ and $q$ at $\rho < 1$ or $\rho
> 1$ (i) the motion of the front is the {\it universal} function
of the "distance" to the collapse time $t_{c}-t$ with the
remarkable property $x_{f}^{<}(t_{c}-t)=1-x_{f}^{>}(t_{c}-t)$.
Moreover, the front velocity $v_{f}$ is the {\it unique} function
of $x_{f}$ with the remarkable symmetry $x_{f}\leftrightarrow
1-x_{f}, v_{f}\leftrightarrow -v_{f}$; (ii) the reduced particle
number $ N_{A}/|\Delta|$ and the reduced boundary current
$J/|\Delta|$ are {\it universal} functions of $t_{c}-t$ with the
remarkable properties $N_{A}^{<}(t_{c}-t)=
N_{A}^{>}(t_{c}-t)-|\Delta|$ and $J^{<}(t_{c}-t)=J^{>}(t_{c}-t)$.

Introducing the relative time ${\cal T}=t_{c}-t$, from Eqs.(13)-
(18) in the vicinity ${\cal T}\ll 1/\pi^{2}$ of the critical point
$t_{c}$ we come to the universal power laws of self-accelerating
collapse ($|v_{f}|\propto {\cal T}^{-1/2}$)
\begin{eqnarray}
x_{f}^{<},1-x_{f}^{>}=\sqrt{2{\cal T}}+\cdots,
\end{eqnarray}
\begin{eqnarray}
N_{A}^{<}, N_{B}^{>}=(\sqrt{8}/3)\pi^{2}|\Delta|{\cal
T}^{3/2}+\cdots,
\end{eqnarray}
\begin{eqnarray}
J=\sqrt{2}\pi^{2}|\Delta|\sqrt{{\cal T}}+\cdots.
\end{eqnarray}
At large $t_{c}\gg 1/\pi^{2}$ far from the critical point ${\cal
T}> 1/\pi^{2}$ according to Eqs. (13)-(18) there is realized the
intermediate exponential relaxation regime ($|v_{f}|\propto
e^{-\pi^{2}{\cal T}}$)
\begin{eqnarray}
x_{f}^{<,>}=1/2\mp e^{-\pi^{2}{\cal T}}/\pi+\cdots,
\end{eqnarray}
\begin{eqnarray}
N_{A}^{<,>} = (|\Delta|/\pi)e^{\pi^{2}{\cal T}}(1\mp\pi
e^{-\pi^{2}{\cal T}}/2+\cdots),
\end{eqnarray}
\begin{eqnarray}
J=\pi |\Delta|e^{\pi^{2}{\cal T}}(1-e^{-2\pi^{2}{\cal
T}}/2+\cdots),
\end{eqnarray}
which in the limit $t_{c}\to\infty$ $(|{\cal C}|, |\varrho -1|\to
0)$ becomes dominant. Thus, at large $t_{c}\gg 1/\pi^{2}$ the
point $x_{f}\approx 1/2$ ({\it stationary front}) is an
"attractor" of trajectories. Exactly at the critical point
$\rho_{\star}=1$ from Eqs. (9) and (10) we find
$x_{f}^{\star}=1/2$ and obtain
\begin{eqnarray}
N_{\star}/N_{0}=\left(\frac{2}{\pi^{2}}\right)\frac{\sin(\pi
q)}{q(1-q)}e^{-\pi^{2}t} +\cdots,
\end{eqnarray}
\begin{eqnarray}
J_{\star}=2(\sin(\pi q)/q)e^{-\pi^{2}t}+\cdots.
\end{eqnarray}
In order to answer the question of when and how the "attractor"
$x_{f}^{\star}=1/2$ is reached it is necessary to retain the next
term ($n=2$) in the sum (8). With allowance for the first two
terms one can easily obtain
\begin{eqnarray}
x_{f}^{\star}=1/2 - D(q)e^{-3\pi^{2}t}+\cdots
\end{eqnarray}
where $D(q)=(A_{2}/\pi A_{1})=\sin(2\pi q)/2\pi\sin(\pi q)$.
According to (27) at $q=1/2$ the coefficient $D$ reverses its
sign, therefore, as it is to be expected, at $q<1/2$ and $q>1/2$
the front reaches the attractor $x_{f}^{\star}=1/2$ from the left
and the right, respectively. By combining (22) and (27), at small
but finite ${\cal C}$ we have $x_{f}^{<,>}=1/2 - {\cal C}
e^{\pi^{2}t}/\pi - D e^{-3\pi^{2}t}+\cdots$. We thus conclude that
under the condition $D{\cal C}>0$ there arises the turning point
of the front ($v_{f}^{M}=0$) with the coordinates
\begin{eqnarray}
t_{M}=(1/4\pi^{2})\ln(\lambda_{M}|D/{\cal C}|)+\cdots,
\end{eqnarray}
\begin{eqnarray}
x_{f}^{M}=1/2 - m_{M} D|{\cal C}/D|^{3/4}+\cdots,
\end{eqnarray}
where $\lambda_{M}=3\pi, m_{M} = 4/(3\pi)^{3/4}$, whereas at
$D{\cal C}<0$ there arises the inflection point of the front
trajectory $(|v_{f}^{s}|={\rm min}|v_{f}|)$ with the coordinates
$t_{s},x_{f}^{s}$ which are determined by Eqs. (28), (29) with the
coefficients $\lambda_{s}=3\lambda_{M}, m_{s}= 2m_{M}/(3)^{3/4}$.
The analysis presented demonstrates the key points of evolution of
the island-island system at arbitrary $q$ and $r$ with ${\cal
C}(q,r)< 1/e$. Below we will focus on a detailed illustration of
this evolution from the initial island-sea configuration ($q\ll
1$).

A remarkable property of the island-sea configuration $q\ll 1$ is
that at $r\gg 1$ the $\Delta(\rho)=\rho-1$ value and all the
coefficients $A_{n}(\rho)=2\rho$ up to $n\propto 1/q \gg 1$ become
unique functions of $\rho$. Therefore, the system's evolution at
$t\gg q^{2}$ is determined by the sole parameter $\rho$. At
$q^{2}\ll t \ll 1$ we have the scaling IS regime \cite{ship}
\begin{eqnarray}
x_{f}=\sqrt{2t}\ln^{1/2}(\rho^{2}/\pi t), \quad
t_{c}(\rho)=\rho^{2}/\pi
\end{eqnarray}
with $x_{f}^{M}=\rho\sqrt{2/\pi e}$, $t_{M}=\rho^{2}/\pi e$. For
$t> 1/4\pi^{2}$ with allowance for two principal modes ($n=1,2$)
we obtain from (8)
\begin{eqnarray}
x_{f}=(1/\pi)\arccos[G(\rho,t)e^{3\pi^{2}t}/4]
\end{eqnarray}
where $G(\rho,t)= \sqrt{1+8{\cal
C}(\rho)e^{-2\pi^{2}t}+8e^{-6\pi^{2}t}}-1$ and ${\cal
C}(\rho)=(1-\rho)/2\rho$. For the time of collapse $t_{c}(\rho)$
we derive from (8) the general equation for arbitrary $\rho$
\begin{eqnarray}
\sum_{n=1}^{\infty}(\pm 1)^{n}e^{-n^{2}\pi^{2}t_{c}(\rho)}= \pm
|{\cal C}(\rho)|
\end{eqnarray}
whence for the leading terms in accord with (31) we find
\begin{eqnarray}
t_{c}(\rho)=(|\ln|{\cal C}||\pm|{\cal C}|^{3}+\cdots)/\pi^{2}.
\end{eqnarray}
Using small $t$ representations of the series (32), one can easily
show that, with the growing $\rho$, $t_{c}$ initially grows by the
law $t_{c}(\rho) = \rho^{2}(1+4e^{-\pi/\rho^{2}}+\cdots)/\pi$,
then it passes through the critical point
$t_{c}(\rho_{\star})\to\infty$ according to Eq.(33), and finally
at large $\rho$ decays by the law $t_{c}(\rho)\propto 1/\ln\rho$.
From Eqs. (31) and (17) for the starting points $t_{M,s}$ of front
self-acceleration at small $|{\cal C}|$ we find
\begin{eqnarray}
t_{M,s}/t_{c}=1/4+\beta_{M,s}/|\ln |{\cal C}||+\cdots
\end{eqnarray}
with the number of $A$ particles $N_{A}^{M,s}/N_{A0}\propto |{\cal
C}|^{1/4}$ where $\beta_{M}=\beta_{s}/2=\ln 3/4$. Remarkably, the
same as for the scaling IS regime (3), (30) in the vicinity
$|\rho-\rho_{\star}|\ll 1$ the ratio $t_{M}/t_{c}$ reaches the
{\it universal limit} $t_{M}/t_{c}=1/4$. In Fig. 1 are shown the
calculated from (30) and (31) trajectories of the front
$x_{f}(t)$, which illustrate the evolution of the front motion
with the growing $\rho$. It is seen that to $\rho\approx 0.7$ the
death of the island $A$ proceeds in the scaling IS regime (30)
($t_{M}/t_{c}=1/e$), then the $x_{f}(t)$ trajectory begins to
deform, and at small $|\rho-\rho_{\star}|\ll 1$ the regime of the
dominant exponential relaxation (22)-(24), (34)
($t_{M}/t_{c}\approx 1/4$) is reached. After the critical point
$\rho_{\star}=1$ has been crossed, the death of the island $A$ is
superseded by the death of the island $B$, so the front trajectory
becomes monotonous, and the stopping point of the front
$x_{f}^{M}, t_{M}$ ($v_{f}^{M}=0$) "transforms" to the point of
maximal deceleration of the front $x_{f}^{s}, t_{s}$
$(v_{f}^{s}={\rm min}v_{f}\propto |{\cal C}|^{3/4}$) which at
large $\rho$ shifts by the law $1-x_{f}^{s}\propto
1/\sqrt{\ln\rho}$ with $t_{s}\propto 1/\ln\rho$.

One of the key features of the island-island problem is a rapid
growth of the front width $w$ while the islands are dying.
Therefore, to complete the analysis we have to reveal the
applicability limits for the sharp front approximation $\eta =
w/{\rm min}(x_{f}, 1-x_{f})\ll 1$. By substituting (21) into (2)
we obtain for the self-accelerating collapse $\eta\sim ({\cal
T}_{Q}/{\cal T})^{\mu}$ where for the mean-field front $\mu_{\rm
MF}=2/3$ and ${\cal T}_{Q}^{\rm MF}=1/\sqrt{|\Delta| k}$. For
perfect diffusion-controlled 3D reaction $k\sim Dr_{a}$ where
$r_{a}$ is the annihilation radius. Thus, as our $k$ is measured
in the units of $D/L^{2}b_{0}$ \cite{ship} for the dimensionless
$k$ we have $k=r_{a}L^{2}b_{0}$. Substituting here $r_{a}\sim
10^{-8}$ cm, $L\sim 10$ cm and $b_{0}\sim 10^{22} {\rm cm}^{-3}$
we find $k\sim 10^{16}$ and derive ${\cal T}_{Q}^{\rm MF}\sim
10^{-8}/\sqrt{|\Delta|}$ so that for not too small $|\Delta|$
($|\rho-\rho_{\star}|\gg 10^{-8}$) the sharp front is not
destroyed almost down to the point of collapse. Clearly that at
small $|\Delta|\to 0$ the "destruction" of the front has to occur
already at the stage of exponential relaxation (22)-(26).
Substituting (26) into (2) for the exponential relaxation we find
$\eta\sim e^{\nu\pi^{2}(t-t_{Q})}$ where $\nu_{\rm MF}=1/3$ and
$t_{Q}^{\rm MF}=(\ln k)/\pi^{2}$. Substituting here $k\sim
10^{16}$ we obtain $t_{Q}^{\rm MF}\sim 3.7$ and then from (25) we
find $N_{\star}^{\rm MF}(\eta=0.1)/N_{0}\sim 10^{-13}$. The
analogous calculation for the fluctuation 1D front gives $\mu_{\rm
F}= 3/4, {\cal T}_{Q}^{\rm F}\sim 1/(|\Delta| n_{0})^{2/3}$ and
$\nu_{\rm F}=1/2, t_{Q}^{\rm F}= (\ln n_{0})/\pi^{2}$ where
$n_{0}=Lb_{0}$. Substituting here $n_{0}\sim 10^{6}$ we find
${\cal T}_{Q}^{\rm F}\sim 10^{-4}/|\Delta|^{2/3}, t_{Q}^{\rm
F}\sim 1.4$ and $n_{\star}^{\rm F}(\eta=0.1)/n_{0}\sim 10^{-4}$.
We conclude that both for the MF and the fluctuation fronts the
vast majority of the particles die in the sharp front regime,
therefore, the presented theory has a wide applicability scope.

In summary, the evolution of the system island $A$-island $B$ at
the sharp annihilation front $A+B\to 0$ propagation has been first
considered and a rich dynamical picture of its behavior has been
revealed. The presented theory may have a broad spectrum of
applications, e.g. in description of electron-hole luminescence in
quantum wells \cite{rap}, formation of nontrivial Liesegang
patterns \cite{ant}, and so on. Of special interest is the analogy
of the island-island problem with the problem of annihilation on
the catalytic surface of a restricted medium where for unequal
species diffusivities in a recent series of papers \cite{shi2} the
phenomenon of annihilation catastrophe has been discovered. Study
of the much more complicated case of unequal diffusivities and
comparison with the annihilation dynamics on the catalytic surface
is a generic and challenge problem for future.

This research was financially supported by the RFBR through Grant
No 05-03-33143.

\begin{figure}
\caption {(Color online) Evolution of the front trajectories
$x_{f}(t)$ with growing $\rho$, calculated from Eqs. (30)(blue
lines) and (31)(red circles) at $\rho = 0.5, 0.7, 0.9, 0.98, 1,
1.02, 1.1$ and $2$ (from left to right). The region of the scaling
IS regime is shaded.} \label{fig 1}
\end{figure}
\end{document}